\documentclass[letterpaper]{article}
\usepackage{aaai}
\usepackage{times}
\usepackage{helvet}
\usepackage{courier}
\usepackage{float}
\usepackage{aaai}
\newfloat{algorithm}{t}{lop}
\floatname{algorithm}{Algorithm}
\usepackage[noend]{algpseudocode}

\usepackage{amsmath}

\usepackage{graphicx}
\usepackage{epstopdf}
\usepackage{tikz}

\usepackage{verbatim}

\usepackage{caption}
\DeclareCaptionType{copyrightbox}
\usepackage{subcaption}

\usepackage{url}
\usepackage{dblfloatfix}

\DeclareGraphicsExtensions{.ps}
\title{QUOTE: ``Querying'' Users as Oracles in Tag Engines \\
A Semi-Supervised Learning Approach to Personalized Image Tagging}

\author{Amandianeze O. Nwana \and Tsuhan Chen \\
School of Electrical and Computer Engineering \\
Cornell University \\ 
Ithaca, New York, 14853 \\
aon3@cornell.edu ~~~~ tsuhan@ece.cornell.edu}

\begin{document}
\maketitle

\begin{abstract}

One common trend in image tagging research is to focus on visually relevant tags,
and this tends to ignore the personal and social aspect of tags, especially on 
photoblogging websites such as Flickr. Previous work has correctly identified that
many of the tags that users provide on images are not visually relevant (i.e. 
representative of the salient content in the image) and they
go on to treat such tags as noise, ignoring that the users \emph{chose} to provide
those tags over others that could have been more visually relevant. Another 
common assumption about user generated tags for images is that the order of these
tags provides no useful information for the prediction of tags on future images.
This assumption also tends to define usefulness in terms of what is visually
relevant to the image. For general tagging or labeling applications that focus on providing
visual information about image content, these assumptions are reasonable, but
when considering \emph{personalized} image tagging applications, these assumptions
are at best too rigid, ignoring user choice and preferences.

We challenge the aforementioned assumptions, and provide a machine learning 
approach to the problem of \emph{personalized} image tagging with the following
contributions: 1.) We reformulate the personalized image tagging problem as a 
search/retrieval ranking problem, 2.) We leverage the order of tags, which does
not always reflect visual relevance, provided by the user in the past as a cue
to their tag preferences, similar to click data,  3.) We propose a technique to
augment sparse user tag data (semi-supervision), and 4.) We demonstrate the
efficacy of our method on a subset of Flickr images, showing improvement over
previous state-of-art methods.

\end{abstract}

%%%%%%%%%%%%%%%%%%%%%%%% TODO %%%%%%%%%%%%%%%%%%%%%%%%%%%%%%%%%%%%
% % A category with the (minimum) three required fields
% \category{H.4}{Information Systems Applications}{General}
% \category{H.1.2}{User/Machine Systems}[Human Information Processing]
% %A category including the fourth, optional field follows...
% %\category{H.3.1}{Information Storage and Retrieval}{Content Analysis and Indexing}
% %A category including the fourth, optional field follows...
% \category{H.3.3}{Information Storage and Retrieval}{Information Search and Retrieval}[Information filtering, Selection process]
% %A category including the fourth, optional field follows...
% %\category{H.3.2}{Information Storage and Retrieval}{Information Storage}[File organization]
% %A category including the fourth, optional field follows...
% \category{I.5.4}{Pattern Recognition}{Applications}[Computer vision]
% 
% \terms{Information Filtering, Computer vision}
% 
% \keywords{tagging, behavior, personalization}

\section{Introduction}
\label{sect:intro}

\begin{figure}[t!]
	\centering
  \includegraphics[width=.4\textwidth]{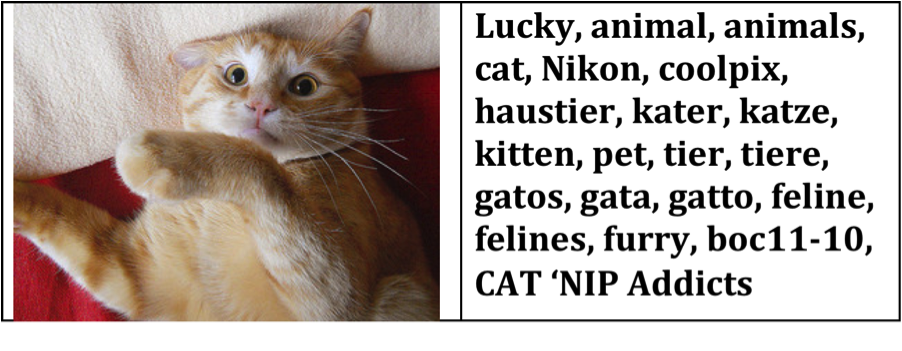}
  \label{fig:flickr_example}
  \caption{
  This example\footnotemark~shows an exemplar image which indicates that the order the
  user presents their tags does not indicate the object of prominence or 
  what is salient, but rather is according to the users judgment of what
  they find relevant. For example, the first tag ``Lucky'' is the name of
  the cat, and is more important to the user than the object level tag ``cat''. }
\end{figure}

\begin{comment}
We aim to challenge the dominant notion of tag usefulness, especially under the
premise of automatic personalized image tagging. The current notion assumes that
tags should be useful as query terms to the image they accompany, and thereby
for the most part only tags that have an unambiguous visual interpretation in
an image are considered.
\end{comment}

The current dominant notion of tag usefulness is that tags should be useful as
search query terms to the images that they accompany, and thereby for the most
part, only tags that have an unambiguous visual interpretation in an image are
generally considered useful in tagging systems. Resulting from this dominant notion of tag 
usefulness, a trend in many papers on image tagging is that image tagging is treated
as essentially a multi-class labeling problem, where the classes
are typically restricted to objects that are salient and visually recognizable
%(cite[tagging as classification])
in the image~\cite{gong:13,li:11,lin:13,rubenstein:12,tomasik:09,wang:11,wray:10}. 
Another trend instead focuses on occurrence and co-occurrence statistics of the tags,
focusing on either frequency or distinctiveness of the tags~\cite{li:08,zhu:10}, 
but even research under this trend operates under the same notion of tag usefulness, 
propagating for the most part only tags that have precise and unambiguous 
visual interpretations. %(e.g. categories in word-net). 

We aim to challenge this dominant notion of tag usefulness, especially under the
premise of personalized image tagging. It has been shown that the 
tags that users provide for their images are categorically different from web
search query terms~\cite{chung:09}.
%(cite:  An Analysis of "Flickr" Tags and Web Image Search Queries - http://eric.ed.gov/?id=EJ869362 )
Hence, we believe that when designing user-centric tagging and indexing systems,
we need to modify the current notion of tag usefulness (which hinges on their
appropriateness as web search query terms) to notions that capture user
preferences and behavior. To do this we have to learn what these user
preferences are, and it is our assumption, as demonstrated by Nwana and Chen~\shortcite{nwana:15}, 
that these preferences can be inferred
from the tag lists users have provided on other images in two ways:
1.) the tags that appear on the list are preferred to those that do not, and
2.) the order that the tags are listed implies a preference on the listed tags
such that those listed earlier are preferred to those listed later. 

The findings of Nwana and Chen~\shortcite{nwana:15} along with a case of a
Flickr\footnotemark~design
change and subsequent reversal mentioned in their paper, demonstrate our
second aforementioned assumption, leading us to believe user tag preferences
can be exploited via tag order.
\footnotetext[1]{https://www.flickr.com/photos/gitpix1/5104408696/in/set-72157624439716281 
| Last Accessed: 01/08/2016}
% \footnotetext[2]{http://www.flickr.com | Last Accessed:01/08/2016}
\footnotetext[2]{https://www.flickr.com/help/forum/en-us/72157645219834187 | Last Accessed:01/08/2016}

% A recent attempt at a design change on Flickr\footnote{http://www.flickr.com Last Accessed:
% 12/16/2015}, and the 
% subsequent reversal of the change, demonstrates our second assumption. The Flickr
% designers opted to update the site to present user generated tags in 
% reverse-chronological order, and immediately active Flickr users protested
% this change, citing that the order that they presented their tags was
% intentional\footnote{https://www.flickr.com/help/forum/en-us/72157645219834187/ Last Accessed:
% 12/16/2015}, leading to an apology by the designers
% and a reverting back to the original chronological order design. Inspired by
% this event, we conducted a study to explore whether users indeed provide
% tags in an order that reflects their preferences even when not prompted to
% do so, and we found that it is indeed the case~\cite{nwana:15}.
% \footnotetext[1]{https://www.flickr.com/photos/gitpix1/5104408696/in/set-72157624439716281 
% Last Accessed: 12/16/2015}

We propose a new method of image tag prediction using learning to rank framework
\cite{hang:11,joachims:02} that attempts to learn the tag ranking functions
which we assume as inherent to each user, so that given a set of good candidate
tags for an image, we can rank them in a manner that best mimics what the user
would have done if limited to that vocabulary. In addition we propose a 
semi-supervised framework to generate more seed tags since user generated tags are typically sparse~\cite{ma:10}
which makes it hard for learning to rank algorithms such as RankSVM~\cite{joachims:02} 
%[to cite] Bridging the Semantic Gap Between Image Contents and Tags
to learn suitable ranking functions. This new prediction paradigm treats the
order that users presented tags in the past as clickthrough data which in the
search engine domain is used as useful implicit user feedback. In this sense
the user ordered tag lists are treated as (noisy and incomplete) oracles.

%----

\subsection{Related Work}
\label{sec:related}

% A popular setting of the image tagging problem is: given an image and a 
% set of initial tags, expand the initial set of tags
% \cite{sigurbjornsson:08,garg:08,wu:09}. Most works under this setting exploits tag
% co-occurrence or correlations to make predictions based on the initial set
% of tags~\cite{menezes:10,belem:11}. Another setting of this problem
% excludes an initial tag set from which to expand~\cite{rendle:10,lipczak:11},
% but instead focuses on tag statistics and image content correlations \cite{wu:09}. 
% There has been some work that has tried to embed both visual features
% and tags into a single multidimensional space in to try to capture the semantic
% correlation between tags and images by their metric distance from one-another
% in that space~\cite{weston:11}. Similarly, there have been works along the lines
% of collaborative filtering and recommendation
% \cite{adomavicius:05,sahoo:12,sen:09,song:08,koren:09} where the task is to predict the missing values or ratings
% of tag to items or $<$user,item/image$>$ pairs. The ``ratings'' in this setting would be inherently
% multidimensional, for example using the bag-of-words representation of the tag list.

Most works in image tagging exploit ``tag-tag'' co-occurrence or correlations to make
predictions based either on an initial seed of tags~\cite{menezes:10,belem:11}
or on ``tag-image content'' correlations and statistics~\cite{rendle:10,lipczak:11,wu:09}.
Some work has tried to embed both visual features and tags into a shared multidimensional
space to try to capture the semantic similarity between tags and images via their distance
from one another in the space~\cite{adomavicius:05,sen:09,song:08,weston:11}. These efforts
are agnostic to user preferences among tags, but rather emphasize the image's ``preferences''
among tags or better yet, a tag's appropriateness to a given image.

Along the lines of orderings and importance of objects in images, Spain and Perona
\shortcite{spain:11} define the importance of an object as the probability that the object
is mentioned first in the tag list for the image given multiple instances of the 
tag lists from \emph{different} users. They assume that each instance of the tag lists
are independently and identically distributed, and hence they learn a global 
notion of importance, while ignoring notions of user importance and preference.
Berg et al.~\shortcite{berg:12} include the notion of object ontologies
to their definition of object importance, with a focus on attribute detection
and scene understanding to determine the importance of objects and appropriateness of tags in
images. But they too focus on global notions of importance, not being able to
capture personal preferences of individual users under their framework. 
Closer to our approach, Hwang and Grauman \shortcite{hwang:12} consider an object's 
importance in the scene as directly proportional to its likelihood of being mentioned early by a
human describing the image. They assume that users mention objects of prominence
in the image early on their tag list, but their application is limited to using object names
as cues for image retrieval applications.

Considering personalization, Rendle and Schmidt-Thieme~\shortcite{rendle:10} represent the tag recommendation
problem under the framework of tensor factorization with the constraints on their learning
objective being for a given $<$user,item/image$>$ pair in the training data, 
the tags mentioned for that pair must be preferred to those that were not, similar to our first assumption, 
but they do not try to learn or enforce a relation between tags that do appear together,
thereby ignoring the structure inherent to the user provided tag lists,
which is our other assumption. Lipczak et al.~\shortcite{lipczak:11} also treat
tags as essentially structureless entities (bag of words) and learn for each user, how
to merge tag statistics for each tag from various modalities. Also, Li et al.~\shortcite{li:11}
in a similar spirit to Lipczak~\shortcite{lipczak:11}, propose a method for learning how to weight
tag scores from multiple tagging functions in order to maximize some desired metric 
(e.g., mean average precision) on training data per user. To our knowledge, no other work
on personalization treats the user provided tag lists as anything more than an orderless,
structureless set, and this we believe is one of the novelties of this work with respect
to others on personalization. And as mentioned earlier, the work by Nwana and Chen~\shortcite{nwana:15} provides
evidence to support that the user tag lists are more than just a bag-of-words.
Our proposed method is different from all the prior in that we learn each user's inherent
tagging functions in a semi-supervised manner based on previously observed rankings by the user.

%% Anonymously cite the toposort work

There have been few other works on learning to rank methods for tagging~\cite{belem:11,canuto:13,wu:09}.
Belem et al.~\shortcite{belem:11} use learning to rank methods,
Genetic Programming, and RankSVM to learn functions that capture how well a candidate
tag describes the content of the object being tagged. There again, we see that tag relevance
is treated in terms of descriptive power of the tag, and its discriminative power among
other descriptive tags. In the work by Lan and Mori~\shortcite{lan:13}, they come up with a
new framework for modeling structured preferences among tags. In their work, their goal is
to rank tags according to their relevance to the image content, and the structure among tags is related
to the object ontologies which the tags describe, rather than the user's inherent preferences. They operate under the
global notion of tag relevance, and their ground-truth ranking of tags
for an image is given by the number of times that the tag is mentioned as visually present in
the image by human annotators on Amazon Mechanical Turk (AMT). 
Wang et al.~\shortcite{wang:10} present a semi-supervised approach
that treats tag lists as inherently unstructured while occasionally querying expert annotators
to rank the tag list according to visual relevance. Liu et al. \shortcite{liu:09} also treat the
tag lists that accompany images as inherently structureless, and their method for tagging
images is purely content based, ignoring higher level semantics and user preferences, and their ground-truth
is the ``majority vote'' of 5 users on each tag into 5 relevance levels (strongly relevant-
strongly irrelevant). They propose a random walk over a tag graph (generated using 
concurrence similarity between tags) to refine tag relevance. Our method is different
from these in that we are concerned with personalized notions of relevance/preference, and we
use the \emph{user provided} tag lists as ground-truth to the correct ranking of tags for that user,
thereby implicitly ``querying'' the user. 

%Talk about other max margin methods? 
%
%On ranking tags: tag ranking (random-walk based rank/score refinement)
%
%On Tag relevance: tag_ranking, grauman, 
%--dog name is probably preferred to ``dog'' for this user.... (fig 1 in tag ranking)
%
%On Importance of objects? spain, berg, grauman, 

%%

%-------------------------------------------------------------------------

\begin{figure*}[t!]
 \centering
  \includegraphics[height=.265\textheight,width=.9\textwidth]{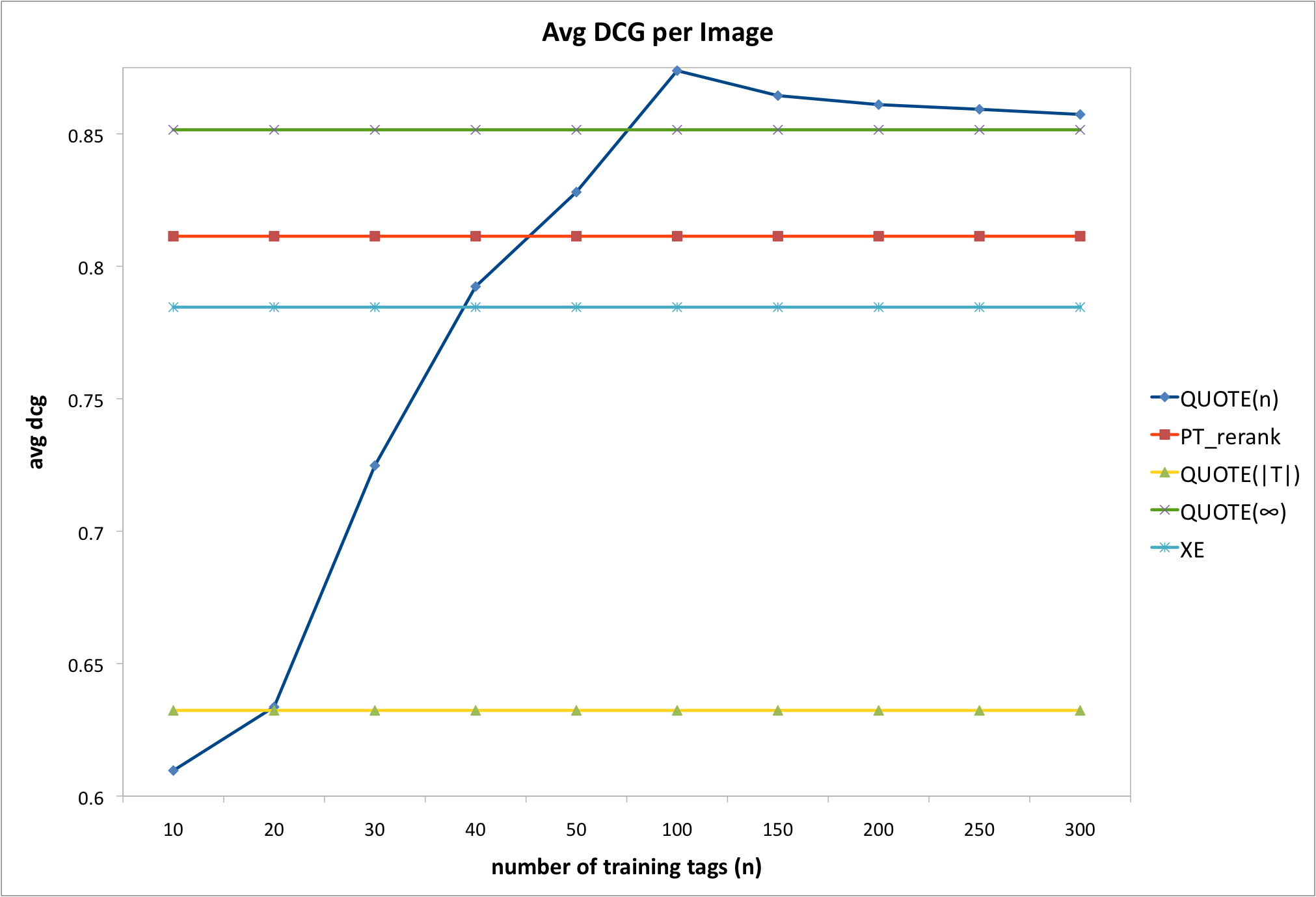}
  \caption{Figure shows how the average $DCG$ changes as a function of the number of
  tags used per training image. The average is taken over all the images in the test
  set. We can see that increasing the number of tags increases the performs up to
  a point where it then saturates. }
  \label{fig:avg_dcg}
 \end{figure*}

%----

\begin{figure*}[t!]
 \centering
  \includegraphics[height=.265\textheight,width=.9\textwidth]{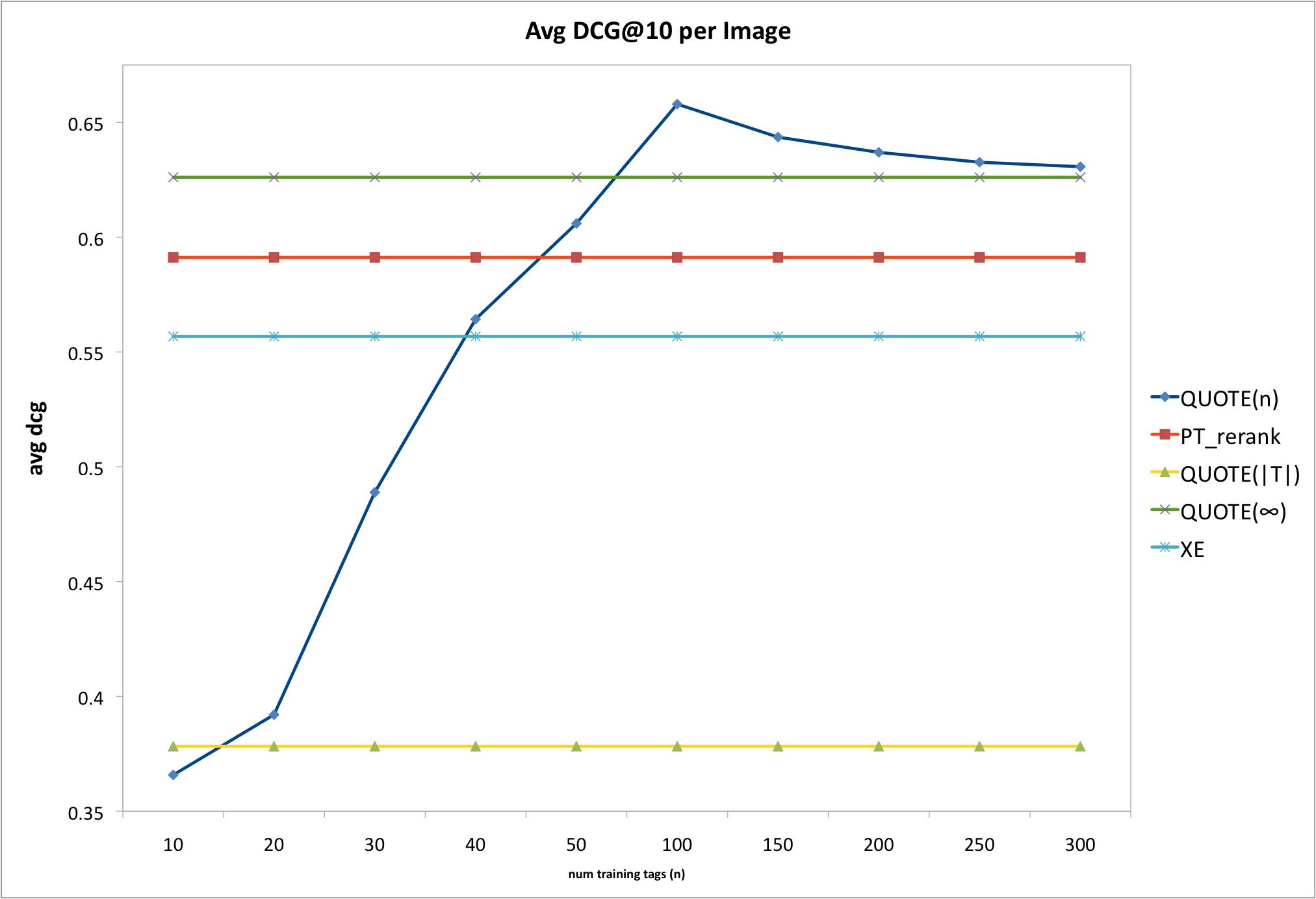}
  \caption{Figure shows how the average $DCG@10$ changes as a function of the number of
  tags used per training image. The average is taken over all the images in the test
  set. We can see that increasing the number of tags increases the performs up to
  a point where it then saturates.}
  \label{fig:avg_dcg_10}
 \end{figure*}

%----

\section{Problem Formulation}
\label{sec:formulation}

We model the problem of personalized automatic image tagging as an instance of
search/retrieval ranking problem, and we employ the learning
to rank solution, RankSVM, by Joachims~\shortcite{joachims:02} to learn our user
models (i.e., ranking functions).

%----

\subsection{General Search/Retrieval Ranking Problem}
\label{sec:ranking_problem}

For a typical search engine, the search/retrieval ranking problem is as follows:
Given a text query, $q$, the system should retrieve a set of related documents,
$\mathcal{D}_q = \{d_1,d_2,\dots\}$, and return, as a result of this query,
the documents in $\mathcal{D}_q$, in order of decreasing relevance. 
% When personalization is considered, the meaning of relevance is no longer considered
% as global, but inherent to the user. 
In the context of personalization, relevance has to be considered as user-specific
rather than global. In most search engine systems,
there is no direct way of learning the user preferences so implicit cues like
clickthrough data on the ranked documents are used as signals of user preferences.
The idea being that returned documents that are clicked are preferred to those
that are not, so future query results are biased towards these implicit signals.
Although clickthrough data may be noisy, and are not perfect relevance judgments,
it has been shown to convey useful information that has been used to improve and
optimize search engines~\cite{joachims:02}.

%----

\subsubsection{Learning to Rank using RankSVM}
\label{sec:svm_rank}

We briefly discuss how RankSVM~\cite{joachims:02} learns ranking functions 
from preference judgments. We define for simplicity a search session, $S$, as 
the tuple of $(q, \mathcal{D}_q, \mathcal{C})$, where $q$ is the query, 
$\mathcal{D}_q$, the set of retrieved documents, and $\mathcal{C} \subset 
\mathcal{D}_q$, the set of clicked documents. From each session, $S$, we derive
pairwise preference judgments, $P_S$, of the form $P_S = \{ d_i \succ d_j : 
d_i \in \mathcal{C}, d_j \in \mathcal{D}_q\setminus\mathcal{C} \}$. We use 
$(d_i,d_j)$ as shorthand for $d_i \succ d_j$ in the rest of the paper. Let 
$\mathcal{S}$ be the training set of all observed sessions. The objective of 
RankSVM simply stated is, given $\mathcal{S}$, learn a ranking function 
$\vec{w}$ that minimizes the number of reversed preference judgments over all
observed sessions. More concretely: 

\begin{flalign}
& minimize:  V(\vec{w}, \vec{\xi}) = \frac{1}{2}\vec{w}\cdot\vec{w} + 
  C\sum\xi_{i,j,k}\\
&subject~to: ~  \nonumber \\
	&~~~ \forall (d_i,d_j) \in P_{S_1} : \vec{w}\Phi(q_1,d_i) > 
  \vec{w}\Phi(q_1,d_j) + 1 - \xi_{i,j,1} \nonumber \\
  & ~~~ \dots \\
  &~~~ \forall (d_i,d_j) \in P_{S_n} : \vec{w}\Phi(q_n,d_i) > 
  \vec{w}\Phi(q_n,d_j) + 1 - \xi_{i,j,n} \nonumber \\
  &~~~ \forall i\forall j\forall k: \xi_{i,j,k} \geq 0
\end{flalign}

Where $\vec{w}$ is the (linear) ranking function, $\Phi(q,d)$ is a mapping
onto features describing the query and document, $\xi_{i,j,k}$ a slack
variable that allows some degree of error in the learned preference judgment
between documents $i$ and $j$ for session $k$, and $C$ the parameter that 
controls the trade-off between minimizing training error, and generalization
(i.e. reducing over-fitting). 

Given a new query, $q'$, to rank the set of retrieved documents, $\mathcal{D}_{q'}
$, one only has to compute $\vec{w}\cdot\Phi(q',d), \forall d \in \mathcal{D}_{q'}$
and sort in descending order. 
% An advantage of Ranking SVM is that $\vec{w}$ can be
% updated in an online manner as more examples are given. 

%----

\subsection{Tag Ranking Formulation}
\label{sec:tag_ranking}

We adapt the framework of RankSVM to our problem of personalized automatic 
image tagging in the following way: a session $S$ is represented by tuple, $(I,V,T)$
where $I$ is the image being tagged and is analogous to the text query, $q$, in 
section~\ref{sec:ranking_problem}, $V$ the set of available tags (or vocabulary), 
analogous to the documents, and $T$ the tags that were given by the user, analogous
to the clicked pages. 

We make the following assumptions on $V$ and $T$:\\
1.) $T$ has inherent structure in that the tags which are mentioned earlier are
implicitly preferred to those mentioned later. Note that this assumption in section
\ref{sec:ranking_problem} would have been analogous to saying that the pages
which were clicked first are preferred to the later pages, but this assumption
would likely not be a good one in that setting since the user is usually 
estimating based on limited information (title and excerpt) that they will find
the page relevant, or get some utility from it. In fact it is probably the case that the fact they keep
clicking implies they haven't found something satisfactory, and they will stop
either when they find a more satisfactory page or after some upper bound on time spent
searching. But in our setting the users have potentially full information,
that is knowledge of the content of the image (this makes sense since they 
observe the image before tagging), and knowledge of the vocabulary (again
another sensible assumption, especially in light of personalization).\\
2.) The elements in $T$ are also implicitly preferred to the elements in 
$V\setminus T$. This is similar to the assumption in section
\ref{sec:ranking_problem} in that given a vocabulary of tags (or set of 
``documents''), the subset of tags the user mentioned for the image are 
analogous to the subset of ``documents'' the user would have clicked on.\\
3.) $V$ also has structure, in that we can define functions over $V$ that measure
the relevance of the tags in $V$ to the image $I$, so that we can order the set
$V$ for image $I$.

%----

\begin{figure*}[t!]
 \centering
  \includegraphics[height=.265\textheight,width=.9\textwidth]{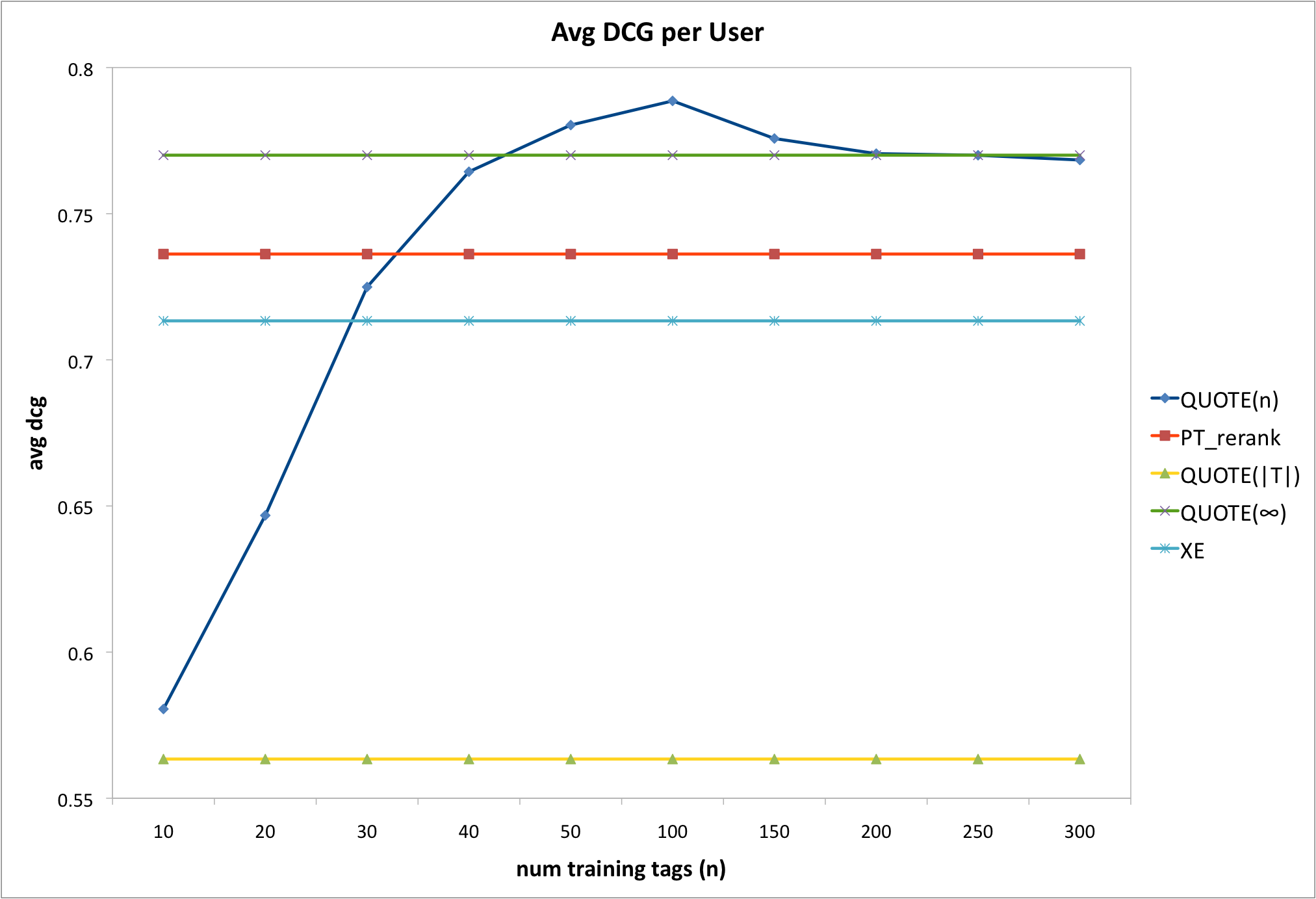}
  \caption{Figure shows how the average $DCG$ changes as a function of the number of
  tags used per training image. The average is taken over the average of the mean 
  performance within each users' images in the test set. We also observe that increasing
  the number of tags increases the performs up to a point where it then saturates.}
  \label{fig:avg_dcg_user}
 \end{figure*}

%----

\begin{figure*}[t!]
 \centering
  \includegraphics[height=.265\textheight,width=.9\textwidth]{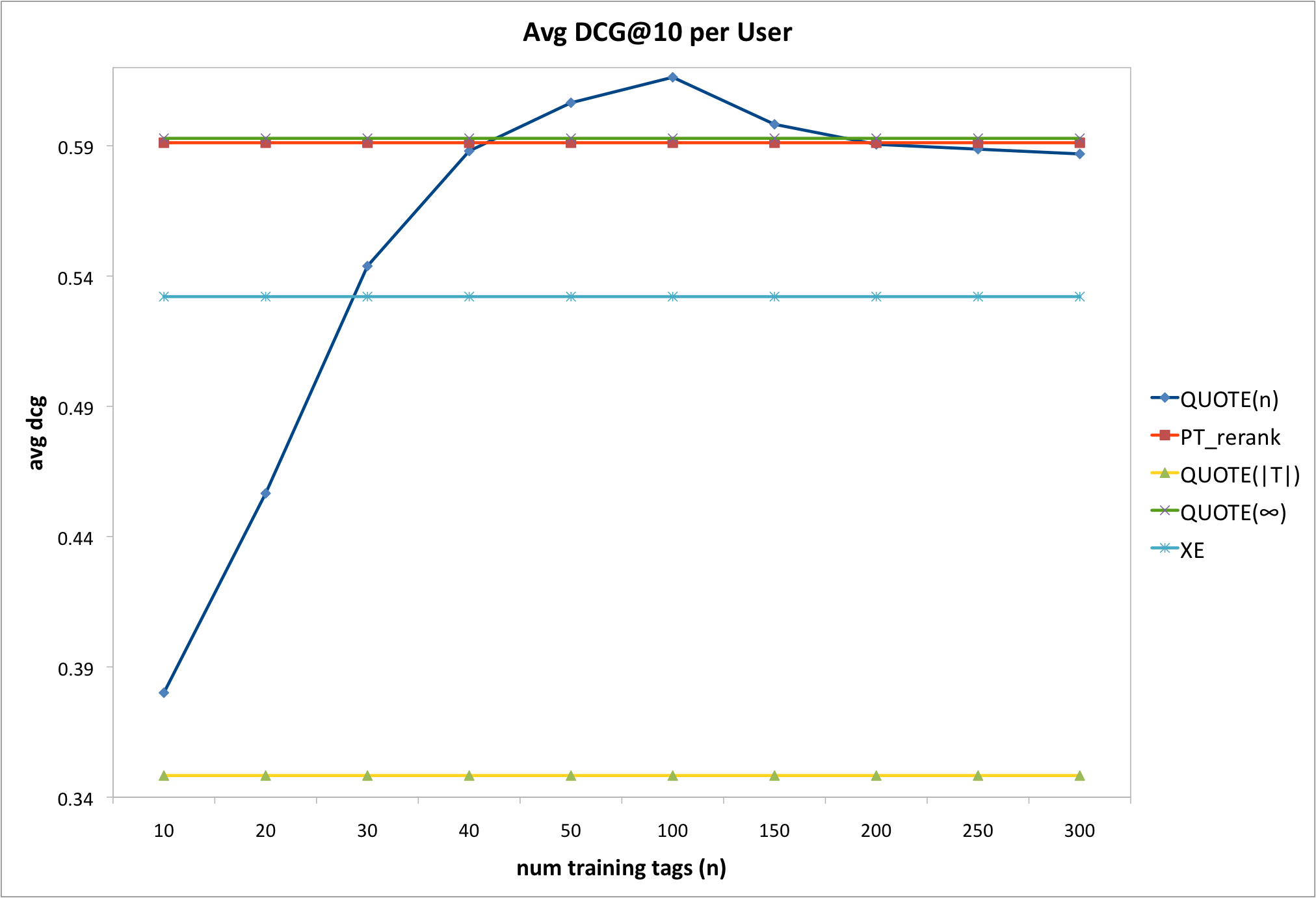}
  \caption{Figure shows how the average $DCG@10$ changes as a function of the number of
  tags used per training image. The average is taken over the average of the mean 
  performance within each users' images in the test set. We also observe that increasing
  the number of tags increases the performs up to a point where it then saturates.}
  \label{fig:avg_dcg_10_user}
 \end{figure*}

%----

\section{Model}

\subsection{Image Representation}
\label{sec:image}

In this work, we follow an implicit (feature similarity based) approach to mining
tags through visually similar images~\cite{li:11,li:08,rendle:10,zhu:10}, in contrast
to explicit (classifier and rule based) approaches. To that
end, every image in our dataset is represented by a 500-D bag of visual-words based
on SIFT \cite{lowe:04} descriptors extracted in training. This allows us to use the euclidean
distance between image descriptors as a measure of their visual similarity. The implicit
approach is preferred because it is more scalable since one doesn't have to learn a
classifier for each new tag word, and also some tag concepts may not be visually representable
which is typically problematic for explicit rule based approaches.

%----

\subsection{Personalization Model}
\label{sec:personal}
We approach personalization in two ways. First, for each user, $u$, we learn a
ranking function $\vec{w}_u$ using their observed ``sessions'', that is, the set
of tuples, $S = (I,V,T), S \in \mathcal{S}_u $ induced by that user. Secondly, for each
user session we generate an ordered set of relevant tags, $\hat{V}(I,u)$, based on the global 
vocabulary, $V$ as follows: Let $T_u(I)$ be the
set of ground-truth tags the user provided for image $I$, and let $\mathcal{I}_u$
be the set of images tagged by the user, we define 
$$pb_u(t) = \frac{|\{I: t \in T_u(I), I \in \mathcal{I}_u\}|}{|\mathcal{I}_u|}$$
as the probability the user, $u$, mentions that tag on any image. Let $\mathcal{I}_{
\Omega}$ be the set of \emph{all images} from \emph{all users}, we define
$$cb(t) = \frac{|\{I: t \in T(I), I \in \mathcal{I}_{\Omega}\}|}{|\mathcal{I}_{\Omega}|}$$
as the probability that any user mentions that tag on any image. Let $NN(I,m)$
be the set of the $m$ most visually similar images to $I$, we define, 
$$sb_{NN(I,m)}(t) = \frac{|\{I': t \in T(I'), I' \in NN(I,m)\}|}{m}$$
as the probability that $t$ is mentioned among the $m$ most similar images to
$I$. Finally we give a score of 
$$v_u(t,I,m) = pb_u(t) + sb_{NN(I,m)}(t) - cb(t)$$
to the tag t, for image $I$. For session $S = (I,V,T)$ in $\mathcal{S}_u$, 
we define, 
\begin{align}
\hat{V}(I,u) = \big< & t_1, t_2, \dots \big> ~~ s.t.~~ t_i, t_j \in V\setminus T, \nonumber \\
 & i\leq j \rightarrow v_u(t_i,I,m) \geq v_u(t_j,I,m) \geq 0
\label{eqtn:v_hat}
\end{align}

Intuitively, the function $v_u(t,I,m)$ measures the uniqueness of the tag $t$
as is suggested by its visually similar neighbors (sb) in light of what is commonly
said (cb) (this is similar in spirit to the TF-IDF score~\cite{book:tfidf}) and is biased toward
what the user has said in the past (pb). This tagging function is similar to the one used
by Li et al.~\cite{li:11} from which greater detail can be gotten. We use the shorthand
$\hat{V}$ for $\hat{V}(I,u)$ where convenient and unambiguous.

So given a user ``session'', $S=(I,V,T)$, let $T$ be represented as an ordered
set $<t_1,t_2,\dots,t_{|T|}>$, under the assumptions from section~\ref{sec:tag_ranking},
our implied preference judgments, $P_S$, becomes:
\begin{equation}
P_S = O_S(T) \cup O_S(T,V) \cup O_S(\hat{V})
\label{eqtn:P_S}
\end{equation}
where $O_S(T)$, $O_S(T,V)$, $O_S(\hat{V})$ are the preference judgments derived from 
the first, second, and third assumptions from section~\ref{sec:tag_ranking} respectively,
\begin{flalign}
\label{eqtn:O_S}
O_S(T) =& \{(t_i,t_j) : t_i,t_j \in T, i < j\} \nonumber \\
O_S(T,V) =& \{(t_i,t_j): t_i \in T, t_j \in V\setminus T\} \\
O_S(\hat{V}) =& \{(t_i,t_j) : t_i,t_j \in \hat{V}, v_u(t_i,I,m) > v_u(t_j,I,m) \} \nonumber
\end{flalign}

With this definition of $P_S$, we are able to learn the personalized ranking
functions in a \emph{semi-supervised} way by augmenting the observed order preferences
in the set $T$ with other relevant unobserved order preferences from $\hat{V}$.

%----

\subsection{Query-Document Mapping}
\label{sec:mapping}

In order to learn a ranking function for the tags, according to section~\ref{sec:svm_rank},
we need to define a mapping, $\Phi(q,t)$, from the query image, and candidate tag to
some feature space. To that end, we use the word2vec
tool\footnote{https://code.google.com/p/word2vec Last~Accessed: 03/24/2015} to learn vector
representations for the tags. We denote the tag $t$'s vector representation from word2vec
as $w2v(t)$. To train our word2vec model, each observed image in our training set
represents a document, with the accompanying tags as the words in the document. We train our word2vec model using
the skip-gram architecture~\cite{mikolov:13,mikolov-nips:13}, and we chose to embed the
tags in 100-dimensional space.

We also included the following tag statistics as features: \\
1.) $mp(t)$: The tag's mean position on the tag lists it appears on. \\
2.) $vp(t)$: The variance of the tag's position on the tag lists it appears on.\\
3.) $cb(t)$: The probability the tag is mentioned on any tag list. 

Finally we have,
$$ \Phi(q,t) = w2v(t) :: mp(t) :: vp(t) :: cb(t) $$
where the $::$ operator is the append operator. You will notice here that the
we do not use the query, $q$, (that is, the image) in our mapping function. We
leave the incorporation of the query to the mapping function as future work. 

%-------------------------------------------------------------------------

% \begin{figure*}[t!]
%  \centering
%   \includegraphics[height=.275\textheight,width=.9\textwidth]{figs/win_rate.png}
%   \caption{Figure shows the win rate of our method with at different numbers of 
%   tags per training image. We see that increasing the number of semi-supervised
%   tags improves not just the absolute average values across our different metrics
%   of evaluation, but also relatively per image and per user our method is preferred
%   more often to the baseline, with $n$, having the same effect as in Figures~\ref{fig:avg_dcg}
%   - \ref{fig:avg_dcg_10_user}}.
%   \label{fig:win_rate}
%  \end{figure*}

 %----

\section{Experiments \& Results}

 %---- 

\begin{table*}[t]
  \begin{center}
    \begin{tabular}{| l || c | c | c | c | c |}
    \hline
              & $QUOTE(|T|)$ & $QUOTE(\infty)$ & $QUOTE(10)$  & $QUOTE(100)$  & $QUOTE(200)$\\
    \hline\hline
    Per Image & -36.0\%      & 5.9\%           & -38.1\%      & \bf{11.3\%}   & 7.7\%   \\
    \hline
    Per User  & -41.1\%      & 0.003\%         & -35.7\%      &  \bf{4.3\%}   & -0.001\%   \\
    \hline
    \end{tabular}
  \end{center}
  \caption{Average DCG@10 percentage improvement over the $PT\_rerank$ baseline. All improvements
  were calculated to have p-value $<$ .001 using a two-sided student's t-test.}
  \label{tab:avg_stats_10}
\end{table*}

%----

\begin{table*}[t]
  \begin{center}
    \begin{tabular}{| l || c | c | c | c | c |}
    \hline
              & $QUOTE(|T|)$ & $QUOTE(\infty)$ & $QUOTE(10)$  & $QUOTE(100)$  & $QUOTE(200)$\\
    \hline\hline
    Per Image & -22.1\%      & 5.0\%           & -24.9\%      &  \bf{7.7\%}   & 6.1\%   \\
    \hline
    Per User  & -23.4\%      & 4.6\%           & -21.1\%      &  \bf{7.1\%}   & 4.7\%   \\
    \hline
    \end{tabular}
  \end{center}
  \caption{Average DCG percentage improvement over the $PT\_rerank$ baseline. All improvements
  were calculated to have p-value $<$ .001 using a two-sided student's t-test.}
  \label{tab:avg_stats}
\end{table*}

 %----

We discuss how we evaluate our model, from 
the choice of the dataset, choice of baseline and choice of evaluation metric. 

\subsection{Dataset}

We work with the NUS-WIDE dataset \cite{nus-wide-civr09} which is a subset of
269,648 images from Flickr. For each image in
the dataset, we know, via the Flickr API, the corresponding user that uploaded
that image, and the sequence of tags that user chose to annotate the image with. 
Since we are particularly concerned about personalization, we only select images
from this database which satisfy the following criteria: the users who uploaded
the image must have at least 6 images in this dataset, similar to the setting in
\cite{li:11}. This results in about 91,400 images from 5000 users. We split this dataset
into a training and test partition, by randomly assigning half of each users' images
to the training, and half to the test set. 
For each image we only retain the tags that occurred frequently enough across
the dataset, in order to make some sort of meaningful inference on the
tags. We made the design choice of working with tags that occurred
at least 50  times in the dataset. This results in a vocabulary of $5,326$ unique tags.

%----

\subsection{Baselines}
\label{exp:baseline}

We use two baselines for comparison, both of which are evaluated on the NUS-WIDE
dataset:

1.)  The first from Li et al. \shortcite{li:11}, which was considered the state-of-art
in personalized image tagging prior to the second baseline method. Their main
idea is that for a given tag, each user has 2 weighting variables, one to weight
how much to rank the tag according to its frequency among the user's past 
images independent of visual content, and the other how much to weight the 
uniqueness of the tag according to its frequency among visually similar images,
versus its frequency from all previous images (not just the user's). These
weights are user dependent, and the score given to a tag is based on the
combination of these two factors.
Their main contribution was a method to optimize these weights. We implement
their method using the two tagging functions corresponding to the two factors
described above, and find the weights for the \textit{PersonalPreference}
\cite{sawant:10} factor and \textit{Visual}\cite{li:10} factor. For the visual 
factor, we also use the 500-D bag of words based on SIFT descriptors for
consistency. They ignore the inherent structure to the tag lists, and treat
tag list as essentially a bag-of-words. We denote this method as $XE$ 
(for cross-entropy).

2.) The second is a heuristic method we proposed (currently under review at another
venue) by demonstrating another use of 
the pairwise tag biases observed from users' past tagging histories to re-rank
existing tag functions. The main idea is that if the pairwise order $<A,B>$ is observed 
significantly more often than $<B,A>$, for a user $u$, then strictly enforce $<A,B>$
in future predictions for that user while preserving pairwise relationships among tag
pairs that need no re-ordering (either because there is no strong preference, or they are
already in the correct order). This method takes the tags from some default rank, $D$
and then re-ranks those tags according to strength of pairwise preferences from the training set.
The pairwise strength is here defined as:

$$p_{ab} = \frac{\text{\# Times tag } a \text{ is mentioned before tag } b}
                {\text{\# Times tags } a \text{ and } b \text{ are both mentioned together}}$$
                
For the purposes of this experiment, we only consider preference strengths
greater than $0.8$ (chosen empirically) so as to prevent over-fitting. These strengths among candidate
tags can be represented succinctly in a directed constraint graph, and a ranking
that respects these constraints can be gotten via a topological sort (using the default
ranking $D$ to resolve ties and cycles). One drawback of this heuristic method is that it is not 
generalizable to tag pairs that never occurred together in data, while the method we propose in this
paper is. We denote this heuristic baseline as $PT\_rerank$.

\subsection{Metrics}
\label{exp:metrics}
Since we assume that the order which a user tags an image is of
some importance, we would like a metric that takes order into
account. And since we are interested in personalization, we treat
the user provided tag list order as the ground-truth order and this is supported
by the claims by Nwana and Chen~\shortcite{nwana:15}.  More concretely, if we have an ordered set of tags, 
$\{t_1,\cdots,t_k\}$, we define the importance or relevance of each tag as follows:
Let $rank^*(t_i)$ be the rank of $t_i$ \emph{with respect to the ground-truth order}
(and $\infty$ if the tag is not among the ground-truth),

\begin{equation}
	rel(t_i) = \frac{1}{rank^*(t_i)},\ \forall t_ i
	\label{eqtn:zipf_rel}
\end{equation}

The above equation is defined as the reciprocal-rank (or zipf-rank), and in 
simple terms states that the $i^{th}$ tag presented by the user
(i.e. ground-truth) is only $\frac{1}{i}$ times as relevant as the first
presented tag. 

More common metrics, such as precision, recall, and average precision, assume
that all tags are equally relevant, so we do not utilize those metrics in this
paper. Instead we use the more appropriate \emph{discounted cumulative gain},
DCG. This is a common metric used in evaluating
search engine results. For an ordered set $T = \{t_1,\cdots,t_k\}$, such that
$i < j \rightarrow t_i \succ t_j$, we define the DCG with respect to the 
ground-truth as:

\begin{equation}
	DCG(T) = rel(t_1) + \sum\limits_{t_i \in T, i\not=1} \frac{rel(t_i)}{\log_2(i)}
	\label{eqtn:dcg}
\end{equation}

Notice that in Equation \ref{eqtn:dcg}, $rank^*(t_i) = i\ \forall i$, \emph{if and
only if} the ranked list $T$ is exactly the same as the ground-truth. This metric
is called \emph{discounted} because the later we include a tag in our ranking,
the less gain we get from it (i.e. its relevance is discounted by the inverse of
the log of its position in the ranking, not the ground-truth).

\begin{comment}
We also define for a given ranked list, $T$, its ideal ranking $\bar{\sigma}(T)$,
such that for $t_i,t_j \in T$, if $i < j$, then $t_i \succ t_j$ in the
ground-truth. Then the normalized DCG is defined as:

\begin{equation}
	nDCG(T) = \dfrac{DCG(T)}{DCG(\bar{\sigma}(T))}
	\label{eqtn:ndcg}
\end{equation}
\end{comment}

Like the precision, recall, and average precision metrics, we can also
%parameterize the DCG, and NDCG to calculate the $DCG@k$ and $nDCG@k$. That is, 
parameterize the DCG metric to calculate the $DCG@k$. That is,
calculate the metric using only the first $k$ entries of the ranked lists. Let
$T[:k]$ be the first k entries of $T$, then:
\begin{equation*}
	DCG@k(T) = DCG(T[:k])
\end{equation*}
\begin{comment}
\begin{equation*}
  nDCG@k(T) = \dfrac{DCG(T[:k])}{DCG(\bar{\sigma}(T)[:k])}
\end{equation*}
\end{comment}

%----

\subsection{Experiment Setup}
\label{exp:setup}

{\bf\underline{Training Phase}}: In the training phase of our experiments, 
for every user, $u$, we produced a ranking example from each image, $i$,
that the user tagged in the training set by collecting the ``supervised''
tag order for that image, $T_i$, as the user provided tag list with order
preserved and the ``semi-supervised'' tag order from tags mined from the
image's most visually similar neighbors and ordered according to 
Equation~\ref{eqtn:v_hat} to produce the semi-supervised set $\hat{V}_i$. 
We then learn the user's ranking function $\vec{w}_u$ by solving the RankSVM objective
using the constraints described in section~\ref{sec:personal}, with the svm-lite
software package.

{\bf\underline{Testing Phase}}: During test time, for a query image, $i$, its
nearest neighbors are used to generate candidate the tags that need to be ranked
by that user's learned ranking function $\vec{w}_u$ from the training phase. 
These candidate tags are the tags from $\hat{V}(I,u)$ according to Equation~\ref{eqtn:v_hat}.

\subsubsection{Parameter Settings and Design Choices}
\label{exp:params}

For our RankSVM formulation we need to make a choice for the regularization
term $C$ which controls the size of the slack variable and hence the trade-off
between training error and generalization. We chose a value of $C=0.01$ in
our experiments. We tried a few choices of $C$ between $0.01$ and $10$ but the
choice did not seem impact the results, so we chose $C=0.01$ since it was the
fastest for training among the values we tried.

Another design decision was how much data to use in creating the
preference orders, $P_S$, per training image as described in section~\ref{sec:personal}.
Given some ``session" $(I,V,T)$, users tend to give too few tags, but using the entire
set $V\setminus T$ as negative tag examples used in creating $O_S(T,V)$ in equation
\ref{eqtn:O_S} might, 1.) slow down the training phase due to too many pairwise
constraints, and 2.) create noisy/meaningless constraints since not all the
tags in our vocabulary are related to the image in question. So to address these
issues we decided to focus not on the whole dictionary but instead on those
tags which are related/relevant as denoted by $\hat{V}$ in equation~\ref{eqtn:v_hat}. 
That is we set $V$ to $\hat{V}$.

We also analyzed the effect of the number of tags given per example by defining,
$\hat{T} = T :: \hat{V}$, which is the tag order defined by equation~\ref{eqtn:v_hat}
(semi-supervised data) appended to the ground-truth user generated tags. To study
the effect of number of tags used we parameterize $\hat{T}(n) = \hat{T}[:n]$, as
the first $n$ elements of $\hat{T}$. With some abuse of notation, this corresponds
to setting $T=\hat{T}(n)$, and $V=\hat{V}=\emptyset$ in equations~\ref{eqtn:P_S} 
and~\ref{eqtn:O_S}.

Since both the training and testing phases of our method require nearest neighbor
search, we also have to decide the number of nearest neighbors to consider, and in 
this work, we chose to use $m=50$ nearest neighbors.

Another design choice was to group the tags in $\hat{T}$ into levels of relevance,
with tags within the same relevance level having no preference among one-another,
but tags in lower levels preferred to tags in higher levels. We assigned the first
5 tags in $\hat{T}$ to levels 1-5 respectively, and then group the next set of 5
tags into increasing levels (For example, tags 6-10 are level 6, and tags 11-15,
level 7). This helps to prevent overfitting due to potential noise in the data,
and also reduces the number of constraints to enforce in training each user,
thereby speeding up training.

 %---- 

\begin{table*}[t]
  \begin{center}
    \begin{tabular}{| l || c | c | c | c | c |}
    \hline
              & $QUOTE(T)$ & $QUOTE(\hat{V})$  & $QUOTE(\infty)$  & $randQUOTE(T)$\\
    \hline\hline
    dcg per image    & 0.632   & 0.817  & 0.852  & 0.455   \\
    \hline
    dcg@10 per image & 0.378 & 0.588  & 0.626   & 0.173  \\
    \hline
    dcg per user     & 0.563  & 0.728  & 0.770  & 0.408  \\
    \hline
    dcg@10 per user  & 0.348 & 0.544  & 0.593  & 0.168  \\
    \hline
    \end{tabular}
  \end{center}
  \caption{This table shows the comparison of the different components of our
  proposed method. $QUOTE(T)$ measures the performance of just the supervised, 
  $QUOTE(\hat{V})$ measures the performance of the non-supervised order, and
  $QUOTE(\infty)$ the combined performance. $randQUOTE$ measures the performance
  not using user-specific models. All improvements were calculated to have p-value $<$ .001
   using a two-sided student's t-test.}
  \label{tab:configs}
\end{table*}
 %----
 
\subsection{Evaluations}
\label{exp:evals}

%-
\begin{figure*}[th!]
 \centering
  \includegraphics[height=.26\textheight,width=.9\textwidth]{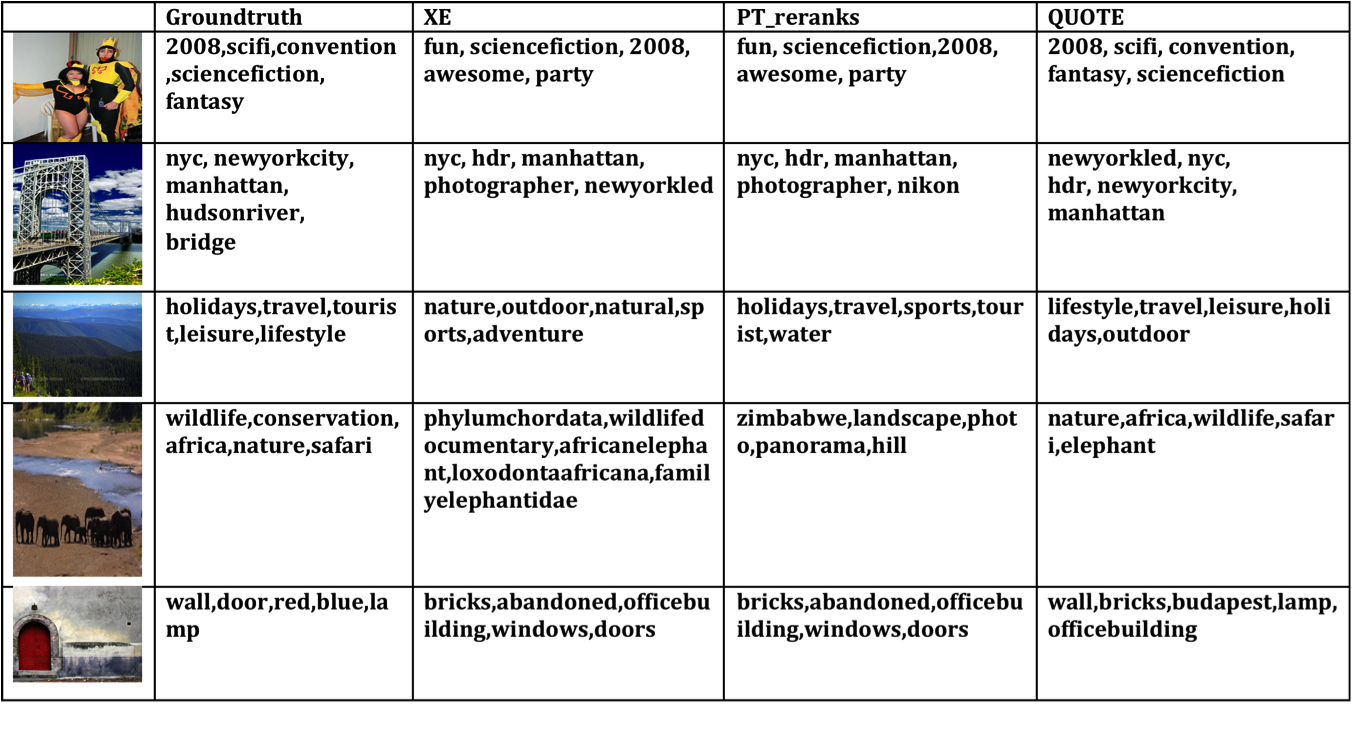}
  \caption{This figure shows example images from our experiment with their
  groundtruth tags and the output of the different algorithms compared in this
  paper.}
  \label{fig:examplar}
 \end{figure*}
 %-

We evaluate our method by using the $DCG$ metric both on a per-image and
per-user basis. Especially since we are concerned with personalization,
we want to know what the average performance of our method is for each
user compared to the baselines. We also present the $DCG@10$ since
in our dataset the average number of tags per-user is about $10$.

We explore the effect of using only supervised orders, $QUOTE(|T|)$, without
supervised orders, $QUOTE(|\hat{V}|)$ and also the effect of using all the tags
generated during the semi-supervised step, $QUOTE(\infty)$. We also explore the
effect of the number of tags, n, per image when training our method, hereafter 
referred to as $QUOTE(n)$. 

% We evaluate our results by the average DCG scores, and also by the win-rate relative
% to our $PT\_rerank$ baseline which already outperforms the current state-of-art
% for automatic personalized image tagging~\cite{li:11}. We define win-rate as:
% $$wr(ALG) = \dfrac{\#ALG \succeq_{DCG} BASELINE}{\#BASELINE \succeq_{DCG} ALG} $$
% where $ALG$ is the method being compared to the baseline.

To make sure that our method truly captures personalization, we also compare
our method to a modified version which in the testing phase uses some other user's
ranking function (randomly) to sort the tags mined from the nearest neighbors. We
refer to this setting as $randQUOTE$.

%----

\subsection{Observations \& Discussion}
\label{exp:observe}

We see from figures~\ref{fig:avg_dcg} - \ref{fig:avg_dcg_10_user} that the number of
ordered tags (clickthrough data) that are provided per training example 
profoundly affects the performance of our method as expected. The more ordered
tags are provided the better performance we get up to a point (100 tags), when
we observe a slight degradation and then saturation. This degradation and 
saturation is probably due to the fact that our method of adding more tags to
the training images in our semi-supervised method start to include some 
less than relevant tags for the image as we provide more tags according to
equation~\ref{eqtn:v_hat}. As we increase more tags past 100, our method
seems to asymptotically approach the performance of when all the semi-supervised
provided tags are included for training. We also notice that at 40 training tags, our
method already begins to match and outperform the $PT\_rerank$ baseline which itself outperforms
the previous state-of-art~\cite{li:11}.Given the size of our vocabulary (over 5000 tags) we do
not think that 40-100 tags per image in training is in anyway excessive, especially
when generated cheaply without need for expert knowledge.

% In figure~\ref{fig:win_rate}, any score greater than 1 implies that the method
% is preferred to the baseline $PT\_rerank$, and we can see that our method, with
% $n=100$ is about 2 times more preferred to the previous state-of-art method. 
The observations under the image level averages (figures~\ref{fig:avg_dcg} and
\ref{fig:avg_dcg_10}), and the user level averages (figures~\ref{fig:avg_dcg_user}
and \ref{fig:avg_dcg_10_user}) are similar. This shows that we do not just learn
global notions of tag preference, but more importantly personalized tag preferences.
So for the ``average'' user in our dataset, we are able to learn a
ranking function that leverages the user's inherent ranking and preference of
tags in order to improve the task of personalized image tagging. Tables
\ref{tab:avg_stats_10} and \ref{tab:avg_stats} show the average percentage
improvement of our method over the $PT\_rerank$ baseline, and we again observe
that using more tags improves our method.

In Table~\ref{tab:configs} we observe that using the constraints without the
supervised order performs better than when we only use the supervised constraint,
and this is not too surprising as we stated earlier, because there are typically
much fewer supervised constraints since users tend to only provide a small
number of tags ($\sim$10 on average). But as expected the combination of both 
the supervised and unsupervised constraints leads to an improvement over both.
We also see that our method indeed learns personal models, since the 
$randQUOTE$ which ranks queries from one user by a random user's ranking function
performs drastically worse.

To verify that the improvements reported in Tables \ref{tab:avg_stats_10},
\ref{tab:avg_stats} and \ref{tab:configs} are statistically significant, 
we performed two-sided student t-tests, and all of our improvement have 
p-values of less than 0.001 which implies that out findings are indeed 
statistically significant. We also provide qualitative examples in Figure~\ref{fig:examplar}.

Our method is also more efficient than the previous state of the art. We use
$O(\#Users)$ parameters, whereas \cite{li:11} requires $O(\#Users\cdot\#Tags)$,
and $PT\_rerank$ requires $O(\#Users\cdot\#{Tags}^2)$ parameters to train the
respective models. For \cite{li:11} the training time per user was 1 minute (not 
including I/O), whereas for ours, it was under 1 second (with I/O time).

%-------------------------------------------------------------------------

\section{Conclusion \& Future Work}

We proposed a novel measurement of tag preferences and a learning to rank framework,
which exploits implicit user preference ``feedback'', and adds tags in a semi-supervised
manner to enrich our data in order to leverage the power of RankSVM.  As opposed to 
prior work we did not make the assumption that tags which are visually relevant
or are about objects in the image are more important to be mentioned before and over
tags that have more higher-level and abstract meanings. Especially for the purpose of
personalization, we believed \emph{that} prevalent assumption to be too restrictive. 

Our experiments demonstrated the efficacy of our approach and contrary to other learning
to rank methods does not require expert knowledge to train nor evaluate, as the 
implicit human signals, used smartly, have been demonstrated to be good enough to
produce state-of-art results for tag personalization. Our method is also more efficient
than the previous state-of-art in tag personalization.

We think it would be an interesting future direction to jointly embed the
image features and tag features to learn a better mapping that leverages
both the image content and tag meaning using deep recurrent and convolutional
neural networks. We also think that it would be interesting to explore what
cognitive processes drive a user's tagging order. Do they follow an ontology in
a top-bottom or bottom-top manner? Is there some latent hierarchy on tags, or
maybe parts-of-speech which exists? What role does sentiment play in tagging?

Another interesting question is could the auto-tagging problem be posed
as an instance of the machine translation problem: Given an image, output the
most likely sequence of words (tags)? We believe that when the restriction to
visually relevant tags is lifted, the personalized image tagging problem 
becomes richer, allowing us to leverage structure and signals in novel ways
as we have demonstrated in this paper.

\bibliography{wsdm16_rank}
\bibliographystyle{aaai}

\end{document}